\documentclass[preprint,aps]{revtex4-1}
\usepackage{amsmath}
\usepackage{amssymb}
\usepackage{graphicx}
\usepackage{epstopdf}
\begin{document}
\title{Synchronization of Coupled Oscillators - Phase Transitions and Entropies}
\author{Steven Yuvan and Martin Bier}
\affiliation{Dept.\ of Physics, East Carolina University, Greenville, NC 27858, USA}
\date{\today}
\begin{abstract}
Over the last half century the liquid-gas phase transition and the magnetization phase transition have come to be well understood.  After an order parameter, $r$, is defined, it can be derived how $r=0$ for $T>T_c$ and how $r \propto (T_c - T)^\gamma$ at lowest order for $T < T_c$.  The value of $\gamma$ appears to not depend on physical details of the system, but very much on dimensionality.  No phase transitions exist for one-dimensional systems.  For systems of four or more dimensions, each unit is interacting with sufficiently many neighbors to warrant a mean-field approach.  The mean-field approximation leads to $\gamma = 1/2$.  In this article we formulate a realistic system of coupled oscillators.  Each oscillator moves forward through a cyclic 1D array of $n$ states and the rate at which an oscillator proceeds from state $i$ to state $i+1$ depends on the populations in states $i+1$ and $i-1$.  We study how the phase transitions occur from a homogeneous distribution over the states to a clustered distribution.  A clustered distribution means that oscillators have synchronized.  We define an order parameter and we find that the critical exponent takes on the mean-field value of 1/2 for any $n$.  However, as the number of states increases, the phase transition occurs for ever smaller values of $T_c$.  We present rigorous mathematics and simple approximations to develop an understanding of the phase transitions in this system.  We explain why and how the critical exponent value of 1/2 is expected to be robust and we discuss a wet-lab experimental setup to substantiate our findings.   
\end{abstract}  
\maketitle

\section{Introduction}

\subsection{The Liquid-Gas Phase Transition}

A small correction to the Ideal Gas Law suffices to capture most of the phenomenology of the liquid-gas phase transition.  In the Van der Waals Equation \cite{MoorePhysChem},
\begin{equation}
\left( P + {a \over v^2} \right) \left( v - b \right) = k_B T,
\label{VanderWaals}
\end{equation}
$P$ denotes the pressure, $v$ is the inverse density (the container volume divided by the number of molecules), $k_B$ is Boltzmann's constant, $T$ is the temperature.  The ``$b$'' takes into account that the involved molecules are not points, but have finite size.  The ``$a$'' is associated with the attractive force between molecules;.  There is a $v^2$ in the denominator for the following reasons. ($i$) For each molecule the amount of interaction with other molecules in proportional to the number of molecules in a specified volume around the molecule and thus to $1/v$.  ($ii$) For the pressure on the wall of the container, the number of molecules in a layer-volume near the wall is what is significant, a quantity that is again proportional to the density, i.e.\ $1/v$.  Effects $(i)$ and $(ii)$ together lead to the $a/v^2$. 

Equation (\ref{VanderWaals}) leads to the isotherms depicted in Fig.\ 1a.  The equation is readily turned into a 3rd order polynomial in $v$.  There are values of $T$ for which there is only one value of $v$ for every $P$.  But for sufficiently small $T$ there are 3 values of $v$ for sufficiently small $P$.  The $T=T_c$ isotherm is the border curve that divides the $P-v$ diagram into the two regions.  The part of the $T<T_c$ curve where $dP/dv >0$ is not realistic.  It was James Clerk Maxwell who realized that, between A and B, the curve has to be replaced by a horizontal line.  It is along this line that liquid and gas coexist and that the phase transition takes place: as volume is increased, pressure stays the same and the system responds by evaporating more liquid. 

\begin{figure}
\centering{\resizebox{14 cm}{!}{\includegraphics{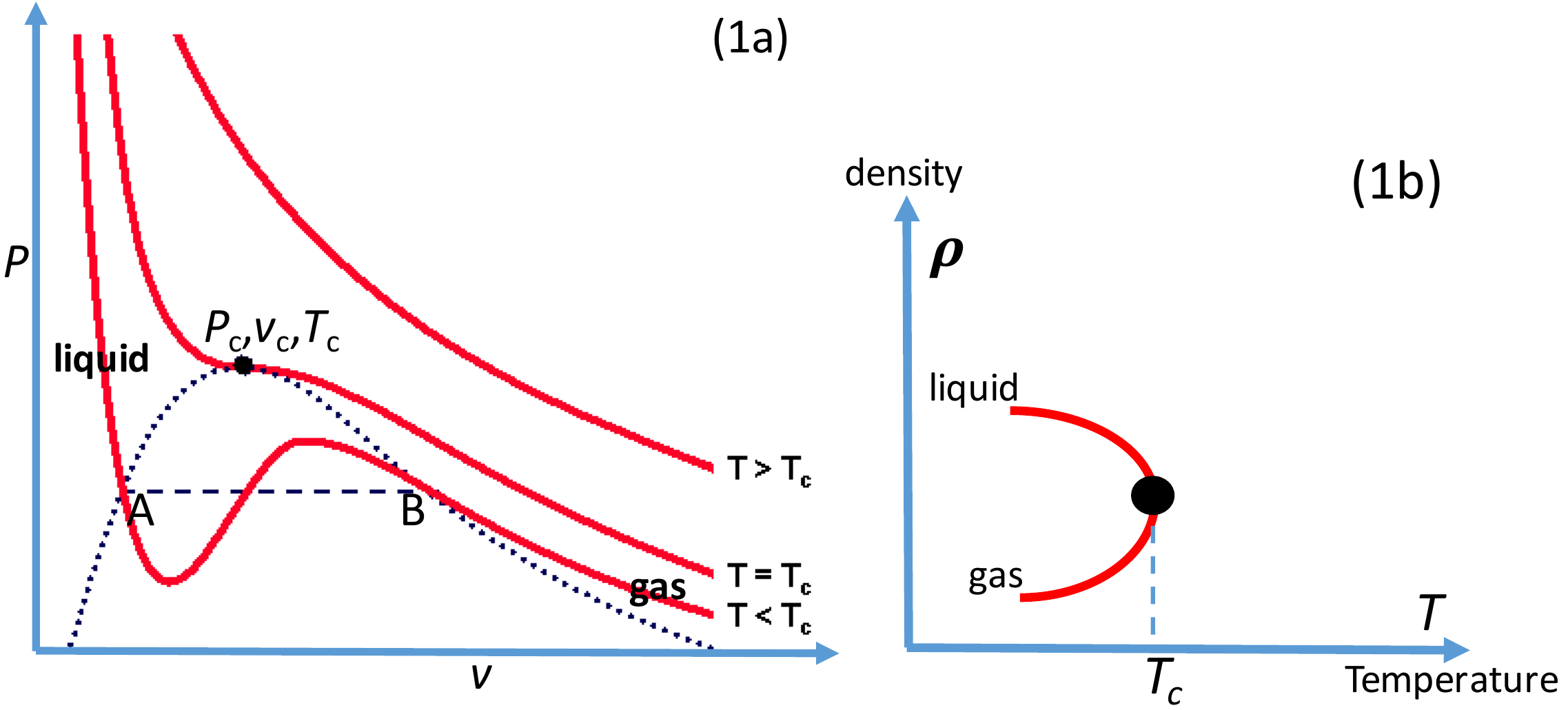}}}
\caption{(a) The $Pv$-diagram according to the Van der Waals Equation (Eq.\ (\ref{VanderWaals})).  In the area inside the dotted curve, there is coexistence of liquid and gas phases.  The critical point $(P_c,v_c,T_c)$ is a unique point where there is no distinction between liquid and gas phase. (b) But lowering the temperature at constant pressure, such distinction emerges and follows a power law $\rho_{liq} - \rho_{gas} \propto \left(T_c - T \right)^{\gamma}$, where $\gamma$ is the critical exponent.}
\end{figure}

On the $T=T_c$ curve there is a point where $d^2P/dv^2 = dP/dv =0$.  This is the critical point.  It is a unique point $(P_c,v_c,T_c)$ in $P,v,T$-space where the liquid phase and the gas phase are not distinguishable.  In the vicinity of the critical point, the different physical quantities follow power laws.  From the Van der Waals Equation it can be derived that at leading order $\left( v_{gas} - v_{liq} \right) \propto \left(T_c - T \right)^{1/2}$ or, equivalently,  
\begin{equation}
\rho_{liq} - \rho_{gas} \propto \left(T_c - T \right)^{1/2}
\label{OrderParamLiqGas}
\end{equation}
In other words, as the temperature is brought from $T_c$ to a small $\Delta T$ below the critical temperature, the liquid and gas densities start to differ proportionally to $\sqrt{\Delta T}$.  This is a symmetry breaking and Fig.\ 1b illustrates how it occurs.  The liquid phase occupies a smaller phase space volume than the gas phase and thus has a lower entropy. The quantity $(\rho_{liq} - \rho_{gas})$, cf.\ Eq.\ (\ref{OrderParamLiqGas}), can therefore be taken as an order parameter.

The critical behavior described by Eq.\ (\ref{OrderParamLiqGas}) does not depend on $a$ or $b$. The exponent $1/2$ is therefore expected to apply universally.  Experiment does show universality.  However, it appears that the actual exponent is not $1/2$, but close to $1/3$ \cite{Guggenheim,ChaikinLubensky}.   

\subsection{Magnetization}   

Magnetization is less omnipresent in daily life than evaporation.  However, as a phase transition it is more easy to model than the transition from liquid to gas.  Figure 2a shows a 2D Ising model.  On each lattice point there is an atom whose spin, $s$, can be pointed either upward $(s=1$) or downward ($s=-1$).  Parallel spins ($\uparrow \uparrow$) have less energy than spins with opposite orientation ($\uparrow \downarrow$).  Assuming that an individual spin only interacts with its four nearest neighbors and that there is no external magnetic field, we have for the magnetic energy of the entire system 
\begin{equation}
H = - J \sum_{\langle i,j \rangle} s_i s_j,
\label{SpinEnergy}
\end{equation}            
where $\langle i,j \rangle$ denotes a summation over all neighbor-neighbor interactions and $-J$ and $J$ are the energies of the parallel and antiparallel orientation, respectively.

\begin{figure}
\centering{\resizebox{12 cm}{!}{\includegraphics{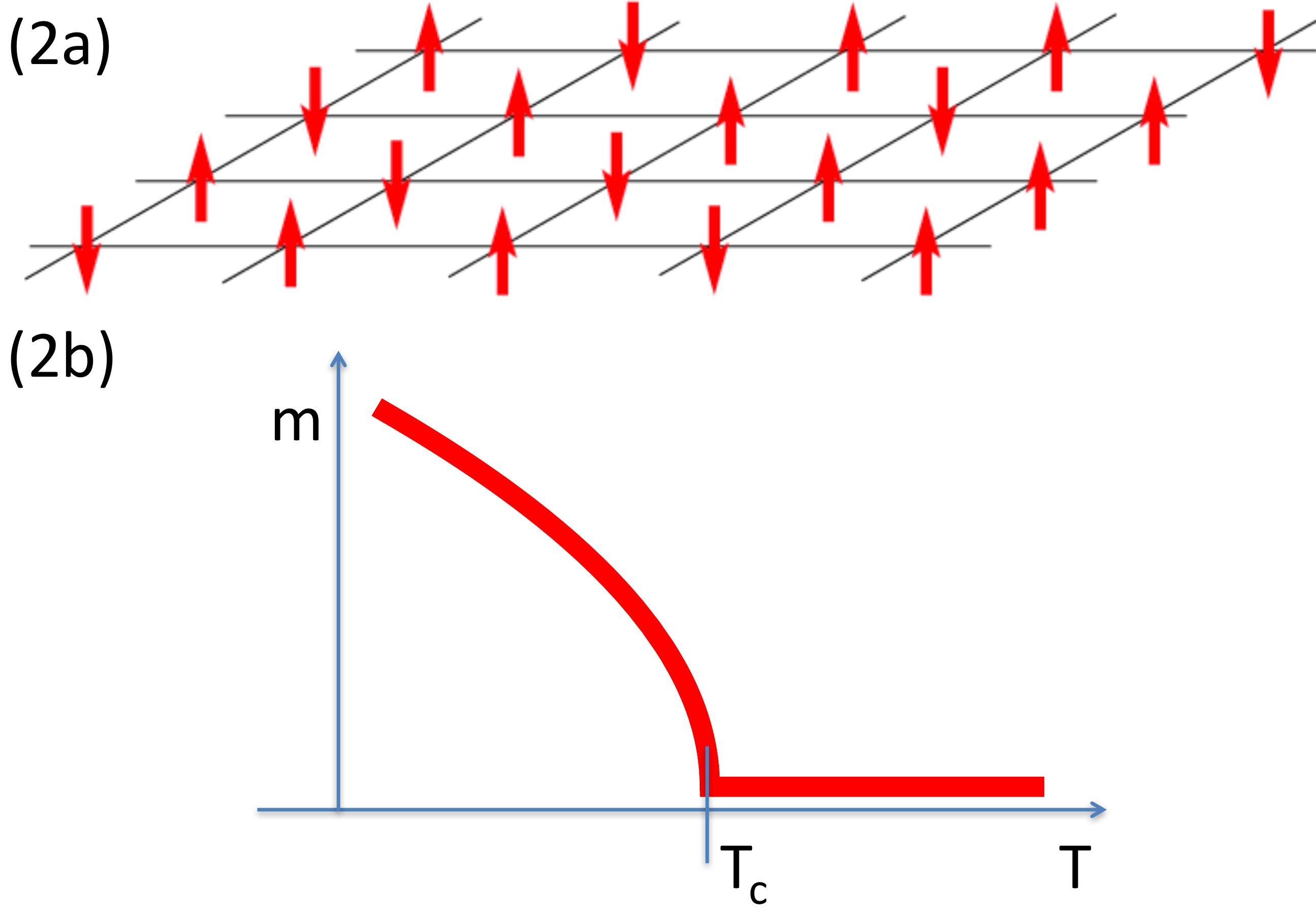}}}
\caption{(a) In the 2D Ising model each spin interacts only with its four nearest neighbors.  (b) Ising models of two and more dimensions exhibit a phase transition.  Below a critical temperature, $T_c$, there is no magnetization.  (b) At the critical temperature there is a discontinuity in the first derivative and below the critical temperature we have $m \propto (T_c - T)^{\gamma}$, where $\gamma$ is the critical exponent.}
\end{figure}

If the temperature is finite, then there is a competition in the system between thermally driven randomization and the ``desire'' of the system to go to the lowest energy by aligning spins.  The solution is readily found if we assume that, through its four neighbors, each individual spin just ``feels'' the average magnetization of the entire system.  This is called the mean-field approximation.  If we let $p$ and $1-p$ be the probabilities of ``spin up'' and ``spin down,'' then we can identify the magnetization of the system, $m$, with the average value of the spin: $m= p(1) + (1-p)(-1)=2p-1$. With the mean-field approximation, the above sum for the energy, Eq.\ (\ref{SpinEnergy}), simplifies: it becomes a sum over all the individual spins and each spin in the system can either be parallel or antiparallel with $m$.  The energy difference between the parallel and the antiparallel orientation is $2 J m$.  The probability $p$ of a spin being parallel to $m$ is then given by a Boltzmann ratio: $p/(1-p) = \exp \left[ 2 J m/(k_B T) \right]$.  With $m=2p-1$, we eliminate $p$ and derive that $m = \tanh \left[ Jm/(k_B T) \right]$.  This equation has three solutions for small $T$ and one solution for large $T$.  As in the case of the liquid-gas transition, the critical temperature $T_c$ marks the border between these two domains. In the vicinity of the critical temperature we take $J/(k_B T) = 1 + \varepsilon$.  Expanding the hyperbolic tangent up to third order in $\varepsilon$, we can solve for the magnetization $m$.  It is found that $|m| \propto \varepsilon^{1/2}$ (cf.\ Fig.\ 2b).  It is obvious that $|m|$ can be seen as an order parameter for the system and that a symmetry breaking occurs if the temperature drops below $T_c$.  

As in the case of liquid-gas transition, the estimate of $1/2$ for the critical exponent turns out to be higher than what experiments show.  Real critical exponents associated with the onset of magnetization cover a range between $1/3$ and $1/2$ \cite{Kadanoff,Arajs,RecentCritExp}.          

\subsection{Universality in Phase Transitions}

For both the liquid-gas transition and the magnetization there is a ``competition'' between the system trying to settle to the order associated with the lowest energy and the thermal agitation due to $k_B T$.  A phase transition occurs when the system exhibits a discontinuity upon variation of temperature.  Though the physical details and the basic equations are different for the magnetization and the liquid-gas transition, in both cases we observed that simple theory predicts a value of 1/2 for an exponent that actual experiment finds to be significantly lower. 

The reason for the discrepancy is the same in both cases.  The mean-field approach is not justified in a 3D reality where each particle only interacts with a limited number of other particles.  The mean-field approximation replaces the product of the two spins, $s_i s_j$, in Eq.\ (\ref{SpinEnergy}) with the product of the average spins, i.e.\ $m^2$.  We can actually rewrite $s_i s_j = (m+\delta_i)(m+\delta_j)$, where $\delta$ can only take the values $(1-m)$ and $(-1-m)$.  What the mean field approximation essentially does is neglect all the $\delta_i \delta_j$ products in Eq.\ (\ref{SpinEnergy}).  This will lead to incorrect energies particularly if there are clusters with parallel spins.

Our treatment of the liquid-gas phase transition was fully based on the Van der Waals Equation.  This equation does not acknowledge local density variations.  The mean-field approximation was therefore implicit when the critical exponent of $1/2$ was derived.

Through analytical solutions and approximations, series expansions, and numerical simulations, the critical exponents for the Ising model in different dimensions have been obtained \cite{ChaikinLubensky}.  There is no phase transition at finite temperature in 1D.  In 2D and 3D the critical exponents are $1/8$ and $0.32$, respectively.  It turns out, finally, that in case of more than three dimensions the number of neighbors is sufficiently large for the mean-field approximation to apply and obtain $1/2$ for the critical exponent. 

\subsection{Phase Transitions for Coupled Oscillators}

Already in the 17th century Christiaan Huygens noticed that two clocks, when hanging side-by-side on a wall, will synchronize their ticking over time.  For a modern scientist or engineer it is not hard to understand that such clocks are mechanically coupled through the little shockwave that propagates through the wall after each tick.  A contemporary and animate version of Huygens' clocks occurred when the London Millennium Footbridge across the Thames was opened in June of 2000.  Pedestrians walking across the bridge started to synchronize their stepping which led to the bridge swaying with an amplitude that was unforeseen \cite{MilleniumBridge}.  Setups with $N$ coupled oscillators are commonly modelled with a system due to Kuramoto \cite{Strogatz}:
\begin{equation}
\dot \theta_i = \omega_i + {K \over N} \sum_{j=1}^N (\theta_j - \theta_i).
\label{Kuramoto}
\end{equation}        
Here $\theta$ and $\omega$ represent the phase and innate frequency of each oscillator.  $K$ denotes the coupling strength between the oscillators.  The last term on the right-hand-side describes a force that drives each oscillator towards the average phase.  As each oscillator ``feels'' the average of the other oscillations, it is obvious that this is a mean-field model.  There is no Brownian noise in this model.  The competition here is between the coupling strength and the distribution of the innate frequencies.  For the Kuramoto model it has indeed been derived and observed that a phase transition occurs as $K$ goes up \cite{Strogatz}.  

It is not just mechanical oscillations that synchronize.  Under anaerobic conditions yeast cells turn glucose into ethanol.  Under certain conditions the throughput of the glucose-ethanol metabolic chain will oscillate with a period of about a minute.  One of the metabolites in the chain, acetaldehyde, can freely permeate the cell membrane.  Yeast cells in a suspension thus share the bath's acetaldehyde concentration and this leads to a coupling.  Experiment and mathematical analysis both show how, in the course of several cycles, the yeast cells in a suspension synchronize their oscillations \cite{BierBakkerWesterhoff}. 

\begin{figure}
\centering{\resizebox{15 cm}{!}{\includegraphics{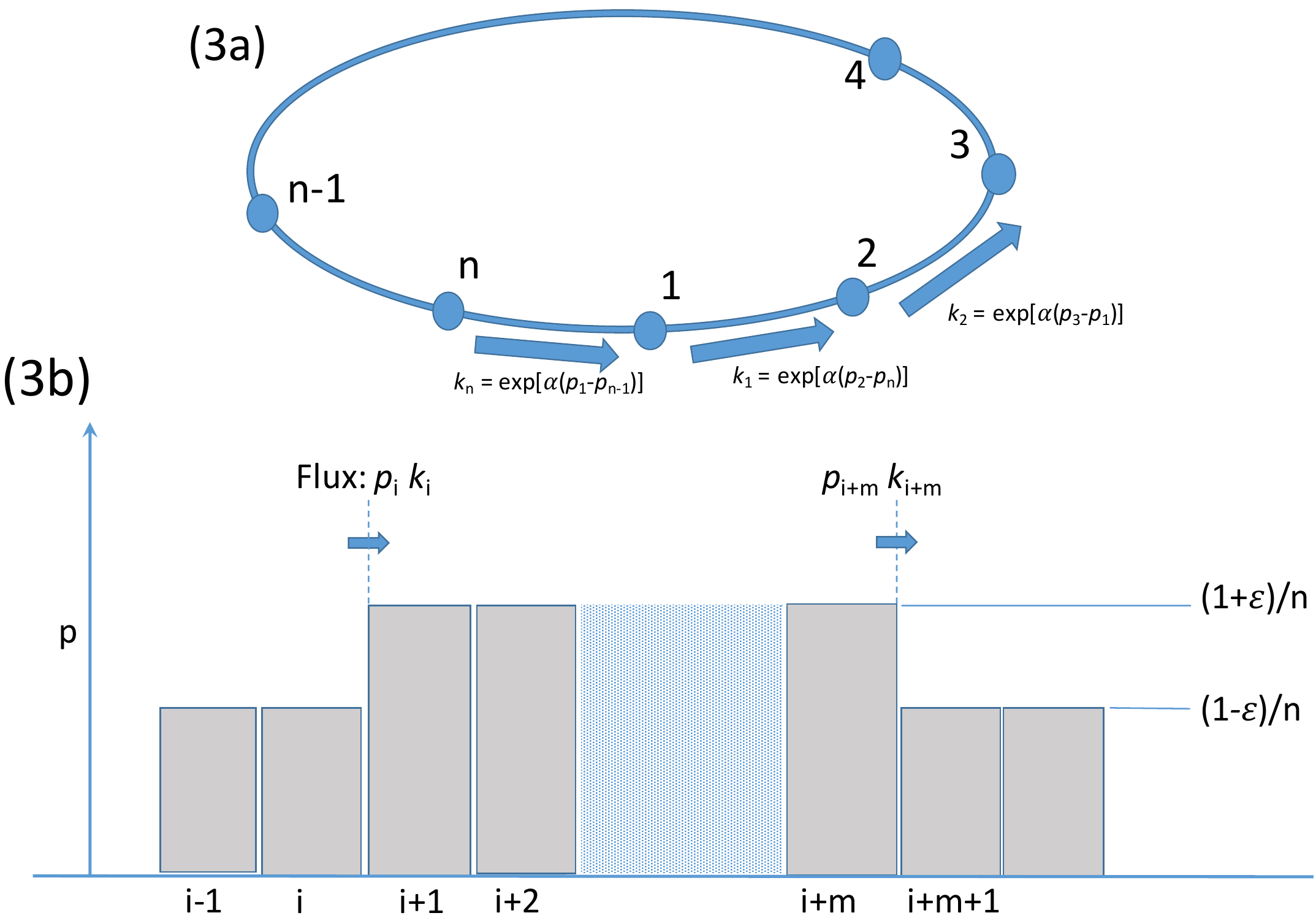}}}
\caption{(a) Our $n$-state model of coupled oscillators.  Each oscillator moves through the $n$-state cycle in the counterclockwise direction.  The transition rate, $k_i$, from state $i$ to state $i+1$ depends on populations in states $i-1$ and $i+1$. (b) For sufficiently low values of the coupling parameter, $\alpha$, there is a homogeneous distribution over the states, i.e.\ $p_i = 1/n \ \forall \ i $.  For values just above the critical value $\alpha_c$, we construct a model in which we assume that the population in a state is either $(1+\varepsilon)/n$ or $(1-\varepsilon)/n$, where $\varepsilon$ is small.}
\end{figure}

A chemically-inspired system of coupled oscillators is what we will focus on in this article. The setup in Fig.\ 3a depicts an $n$-state cycle.  Only counterclockwise transitions are possible.  A large population of oscillators is going through the cycle.  We take $p_i$ to be the fraction of the total population that is in state $i$.  We let the transition rate $k_i$ from state $i$ to $(i+1)$ depend on $p_{i+1}$ and $p_{i-1}$, i.e.\ on the populations in the state ahead and the state behind state $i$ (cf.\ Fig.\ 3a):
\begin{equation}
k_i = k_0 \exp \left[ \alpha \left( p_{i+1} - p_{i-1} \right) \right] .
\label{TransRates}
\end{equation}   
The constant $k_0$ is the same for all transitions and can be absorbed in the timescale; we will leave it out in the remainder of this article.  The idea of Eq.\ (\ref{TransRates}) is that the population in state $(i+1)$ increases the transition rate $k_i$ and thus pulls the population in state $i$ forward to state $(i+1)$.  At the same time, the population in state $(i-1)$ decreases $k_i$ and thus pulls back on the population in state $i$.  For $\alpha > 0$, Eq.\ (\ref{TransRates}) describes a tendency of the entire population to cluster in one or more states.  That tendency increases with $\alpha$.  We choose to put the populations in the exponent as the transition rate, $k$, generally depends exponentially on the height $E$ of the activation barrier associated with the transition, i.e.\ $k \propto \exp \left[ -E \right]$.  In Refs.\ \cite{wood1,wood2,wood3,wood4} a 3-state model with $k_i = k_0 \exp \left[ \alpha \left( p_{i+1} - p_{i} \right) \right]$ is the basis for the analysis, i.e., not the population in the previous state $(i-1)$, but  the population in state $i$ itself is impeding the forward transition from $i$ to $(i+1)$.  

The system in Fig.\ 3a with Eq.\ (\ref{TransRates}) can be taken to model a number of real-life systems.  Ion pumps in cell membranes go through a sequence of conformational states as they go through their catalytic cycle.  An example is the well-known Na,K-ATPase.  This is an ion pump that hydrolyzes ATP and uses the released energy to pump sodium and potassium ions through the membrane against the electrochemical potential \cite{Lauger}.  Na,K-ATPase can be present in the membrane in large concentrations.  These proteins are very polar and they can be coupled through dipole-dipole interaction.  But they can also interact as they change shape in the course of the catalytic cycle and thus deform the cytoskeleton and bilayer membrane.  Through these couplings and the mechanism of Fig.\ 3, the pumps can synchronize their catalytic cycles.

The dynamics of Fig.\ 3a and Eq.\ (\ref{TransRates}) can also be taken to describe the dynamics in a bicycle race.  Due to aerodynamic draft, bicycle racers that ride in a group put in a smaller effort as compared to when they ride alone.  The larger the group, the smaller the power that has to be produced by an individual rider.  Bicycle racers thus tend to cluster in groups where they share the burden.

There is no explicit Brownian noise in the system of Fig.\ 3a.  However, a constant rate out of a state implies that an individual oscillator has an exponentially distributed waiting time in that state.  Effectively, this gives the system a stochasticity and a temperature: for $\alpha = 0$ the $n$-state system will, over time, forget any initial distribution over the $n$ states and evolve towards a homogeneous distribution over the states, i.e.\ $p_i = 1/n$ where $i =1, 2, \ ...\ n$.  The parameter $\alpha$ denotes the coupling strength.  For $\alpha = 0$ there is no coupling.  As in the case of the gas-liquid transition and the magnetization, the randomization is opposed by a coupling that drives the system to a ordered state.   

There is a fundamental difference between the liquid-gas transition (Fig.\ 1) and the magnetization (Fig.\ 2) on the one hand and our setup (Fig.\ 3a) on the other hand.  With the first two systems we are looking at a thermal equilibrium.  The system in Fig.\ 3a, however, goes through a chemical cycle.  As there are no clockwise transitions, there is no detailed balance and there is continuous production of entropy.     

\section{Numerical Simulation and Mathematical Analysis}     
 
\subsection{A Heuristic Approach} 

A simple heuristic model for the behavior close to the phase transition leads to some concise formulae that well approximate the actual behavior.  The simple model can, furthermore, help build intuition for the mechanisms behind the phase transition.  A less general form of the model was also presented in Ref.\ \cite{BierLisowskiNowak} in the context of $n=4$.

As was mentioned before, when $\alpha$ is small, the probability distribution over the $n$ states will over time approach $p_i = 1/n$ for all initial conditions.  For sufficiently large values of $\alpha$, clusters as in Fig.\ 3b will persist.  For the clustering and the homogenization to be in balance, a ``bump'' as in Fig.\ 3b needs to have as much influx ($J_i^{\rm in} = p_i k_i$) as outflux ($J_{i+m}^{\rm out} = p_{i+m} k_{i+m}$).  We are interested in the region near the phase transition, so we consider values of the amplitude $\varepsilon$ (cf.\ Fig.\ 3b) that are small relative to 1.  For simplicity, we allow two values for $p_i$: $(1+\varepsilon)/n$ and $(1-\varepsilon)/n$.  Each value is taken on by half of the probabilities in the distribution.  In the vicinity of the phase transition, equating the fluxes into and out of the ``bump'' leads to:
\begin{equation}
{1-\varepsilon \over n} \exp \left[ 2 \alpha \varepsilon/n \right] \approx {1+\varepsilon \over n} \exp \left[-2 \alpha \varepsilon/n \right].
\label{BierApprox}  
\end{equation}  
This formula is only valid for a cluster that consists of two or more neighboring states with a population $(1+\varepsilon)/n$.  Furthermore, to the left of this cluster there need to be at least two neighboring states with a population $(1-\varepsilon)/n$.  Expressing $\alpha/n$ in terms of $\varepsilon$, we find from Eq.\ (\ref{BierApprox}): $\alpha/n \approx 1/(4 \varepsilon) \log \left[ (1+\varepsilon)/(1-\varepsilon) \right]$.  Expanding the right-hand-side for small $\varepsilon$ up to second order, we next obtain $\alpha/n \approx (1/2) + (1/6) \varepsilon^2$.  From the latter expression, we solve for $\varepsilon$ and thus derive an approximation for the amplitude as a function of the coupling parameter $\alpha$: 
\begin{equation}
\varepsilon \approx \sqrt{6 \over n} \ \sqrt{\alpha - {n \over2}}.
\label{amplitude}
\end{equation}

In order to quantify to what extent a distribution on a cycle as in Fig.\ 3a is homogeneous, an order parameter, $r$, is commonly defined as: 
\begin{equation}
 r e^{i \psi} =  \sum_{k=1}^{n} p_{k} \exp\left[ {2 i \pi k \over n} \right],
 \label{OrderParam}
\end{equation} 
where the ``$i$'' in the numerator of both exponents denotes $\sqrt{-1}$.  This definition is due to Lord Rayleigh \cite{Batschelet,Strogatz}.  It is obvious that we get $r=0$ if there is no synchronization and all the $p_k$'s are identical.  We have $r=1$ if there is maximal synchronization and all molecules are in the same state $j$, i.e.\ $p_k=1$ if $k=j$ and $p_k = 0$ if $k \neq j$.  We consider our model (Fig.\ 3) for a large value of $n$ and we again take a simple approach to come to an upper bound for the value of the order parameter.  Imagine that  $p_k = (1+\varepsilon)/n$ for $1 \leq k \leq n/2$ and $p_k = (1-\varepsilon)/n$ for $n/2 < k \leq n$.  The sum on the right-hand-side of Eq.\ (\ref{OrderParam}) now reduces to $(2 \varepsilon)  \sum_{k=1}^{n/2} (1/n) \exp \left[ i 2 \pi k/n \right]$.  For large $n$ we can approximate the summation with an integral over a half of a period $L$: $(1/L) \int_{x=0}^{L/2} \exp \left[ i 2 \pi x /L \right] \, dx = i/\pi$.  With this result and with Eq.\ (\ref{amplitude}), we derive for the order parameter as a function of $\alpha$:
\begin{equation}
r =0 \ \ {\rm if} \  \  \alpha < n/2 \ \ \ {\rm and} \ \ \  r \approx {3 \over 2} \sqrt{1 \over n} \,  \sqrt{\alpha - n/2} \ \ {\rm if} \ \ \alpha > n/2.
\label{OrderParTheor}
\end{equation}  
This equation makes strong statements about the phase transition.  As the number of states in the cycle, $n$, increases, the phase transition occurs for ever larger values of the coupling parameter as $\alpha_c \approx n/2$.  The critical exponent, however, maintains its mean-field value of 1/2 for all values of $n$.  In the next section we will verify some of these predictions, with both analytical and numerical work.  In the Conclusions and Discussion section of this article we will put these results in the larger phase-transition context.           

\subsection{The System of ODEs}

The system of coupled ordinary differential equations associated with the setup shown in Fig.\ 3a is:
\begin{eqnarray}
\dot p_1 &=& k_n p_n - k_1 p_1, \nonumber \\ 
\dot p_2 &=& k_1 p_1 - k_2 p_2,  \nonumber \\
... \nonumber \\
\dot p_n &=& k_{n-1} p_{n-1} - k_n p_n.
\label{ODEs}
\end{eqnarray}
Here the $k$'s represent the transition rates (cf.\ Eq.\ (\ref{TransRates})) and a dot above a symbol denotes differentiation with respect to time, i.e.\ $^{\bullet} \equiv d/dt$.  The periodic boundary conditions imply $k_n = \exp \left[ \alpha \left( p_1 - p_{n-1} \right) \right]$ and $k_1 = \exp \left[ \alpha \left( p_2 - p_{n} \right) \right]$.  The point $p_j = 1/n \ \forall j$ is the obvious fixed point.  As in an ordinary Taylor series, the behavior of the system in the close vicinity of a point is determined by the lowest order terms, generally the linear terms, in an expansion.  This leads to an $n \times n$ matrix; the so-called Jacobian matrix \cite{BoyceDiPrima}.   The $j$-th row of this matrix lists the values of the derivatives of $(k_{j-1} p_{j-1} - k_j p_j)$ at the fixed point.  The expression $(k_{j-1} p_{j-1} - k_j p_j)$ is the right-hand-side of the $j$-th equation in Eq.\ (\ref{ODEs}).  Along a row of the matrix, derivatives are taken with respect to $p_1, p_2, ... , p_n$, respectively.  We obtain for the Jacobian matrix in our case:
 \begin{gather}
 {\bf J} =
\begin{bmatrix}
	({\alpha \over n} -1)  & -{\alpha \over n}  & 0 &  0  & \dots & 0 &- {\alpha \over n} & ({\alpha \over n}+1) \\
	({\alpha \over n}+1) & ({\alpha \over n}-1)  & -{\alpha \over n} & 0 & 0 &\dots & 0 & - {\alpha \over n} \\
	- {\alpha \over n} & ({\alpha \over n}+1) & ({\alpha \over n}-1)  & -{\alpha \over n} & 0 & \dots & 0 & 0 \\
	0 & - {\alpha \over n} & ({\alpha \over n}+1) & ({\alpha \over n}-1)  & -{\alpha \over n}  & 0 & \dots &  0 \\
	\vdots   & 0 & \ddots  & \ddots & \ddots & \ddots  & 0 & \vdots \\
	\vdots   & \vdots & 0 & \ddots  & \ddots & \ddots & \ddots  & 0 \\
	0 & 0 & \dots & 0  & - {\alpha \over n} & ({\alpha \over n} +1) & ({\alpha \over n}-1)  & -{\alpha \over n}  \\
	-{\alpha \over n} & 0 & 0 & \dots & 0 & - {\alpha \over n} & ({\alpha \over n}+1) & ({\alpha \over n}-1) 
\end{bmatrix} .
\label{Jacob}
\end{gather}
The eigenvalues of this $n \times n$ matrix will tell us whether $p_j = 1/n \ \forall j$ is an attractor or a repeller \cite{BoyceDiPrima}.  For sufficiently small $\alpha$ the real parts of all eigenvalues are negative and the point $p_j = 1/n \ \forall j$ is then an attractor.  The phase transition to synchronization occurs when, upon increasing $\alpha$, the real part of just one eigenvalue turns positive.  At that value of $\alpha$, $p_j = 1/n \ \forall j$ becomes an unstable solution.

The matrix {\bf J} (cf.\ Eq.\ (\ref{Jacob})) is a tetradiagonal circulant matrix.   In a circulant matrix each row is rotated one element to the right relative to the preceding row.  The eigenvalues are the values of $\lambda$ that solve the equation:
\begin{equation}
\det \left( {\bf J} - \lambda {\bf I} \right) = 0.
\end{equation}
A standard formula is available for the eigenvalues of a circulant matrix \cite{Seber}. For the present system with only four possible nonzero elements in the matrix, we obtain after some algebra:

\begin{equation}
\lambda_j = \left( \frac{\alpha}{n}-1 \right) - \frac{\alpha}{n}\exp\left[2\pi i\frac{j}{n}\right] - \frac{\alpha}{n}\exp\left[-4\pi i \frac{j}{n}\right]  + (\frac{\alpha}{n}+1) \exp\left[-2\pi i \frac{j}{n}\right], \ \ j = 0, 1,\ldots,n-1.
\end{equation}
Considering only the real parts, we have:
\begin{equation}
{\rm Re}[\lambda_j] = \frac{\alpha}{n} \left( 1-\cos \left[ 4\pi \frac{j}{n} \right] \right) -1+\cos \left[ 2\pi\frac{j}{n} \right].
\label{RealPartsEigVals}
\end{equation}
Note that $\lambda_0=0$.  This zero eigenvalue is associated with the zero determinant of the above matrix, Eq.\ (\ref{Jacob}), and ultimately with the normalized total population, $\sum_{j=1}^n p_j =1$. It is obvious from Eq.\ (\ref{RealPartsEigVals}) that all eigenvalues for $j \ge 1$ have negative real parts for sufficiently small $\alpha$.  Real parts are zero for $\alpha = (n/4)\sec^2(\pi j/n)$. The smallest values of $\sec^2 (x)$, and hence the first eigenvalues to become positive, occur for arguments furthest away from the asymptote at $x = \pi/2$. This happens simultaneously for $j=1$ and $j=n-1$ and leads to a critical point point at $\alpha_c=(n/4)\sec^2(\pi/n)$.  For large values of $n$, the secant squared will approach unity.  We thus have $\lim_{n \rightarrow \infty} \alpha_c = n/4$.  This is in good agreement with the result of the heuristic approach.

\begin{figure}
\centering{\resizebox{15 cm}{!}{\includegraphics{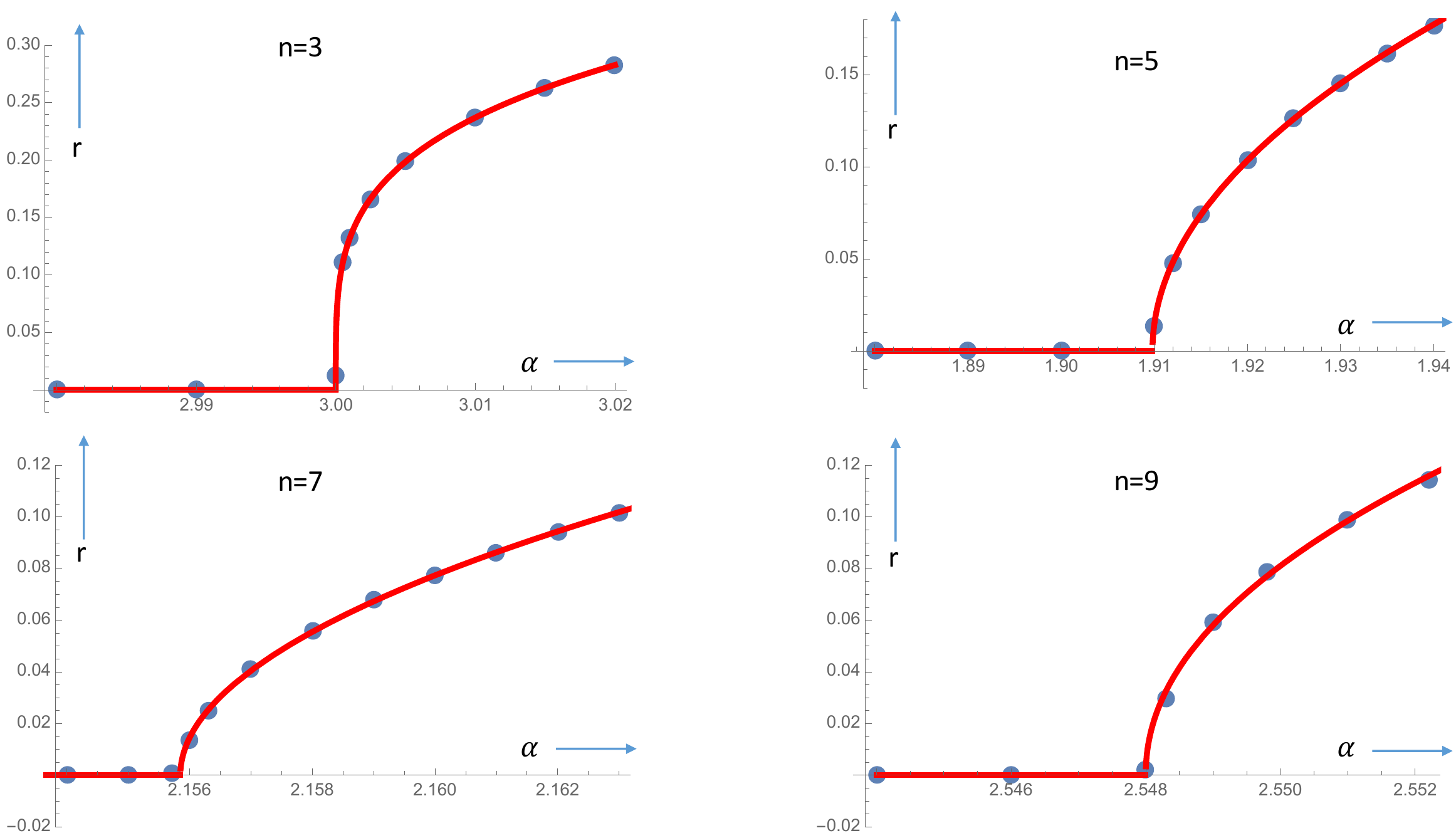}}}
\caption{The order parameter, $r$ (cf.\ Eq.\ (\ref{OrderParam})), as a function of the coupling parameter $\alpha$.  Shown are the results of numercal simulation of the ODEs (cf.\ Eq.\ (\ref{ODEs})) with {\it Mathematica} for 3, 5, 7, and 9 states.  Each dot represents the result of a simulation of a million units of time.  After the order parameter had relaxed to a constant value, the average over 100,000 units of time was taken.  The red curves result from fitting a power law, $r= p (\alpha - \alpha_c)^{\gamma}$ to the blue dots, where $\alpha_c$ is the critical value $\alpha_c=(n/4)\sec^2(\pi/n)$.  Let $(p_n,\gamma_n)$ be the result of this fit for the $n$-state system.  For $n=3$, we find $(p_3,\gamma_3)=(0.77,0.25)$.  For $n= 5$, $n=7$, and $n=9$ we gathered data up to $r \approx 0.1$ so as to identify just the leading order behavior.  The results were $(p_5,\gamma_5)=(1.0,0.50)$, $(p_7,\gamma_7)=(1.2,0.51)$, and $(p_9,\gamma_9)=(1.6,0.48)$.}
\end{figure}

Figure 4 shows the results of numerical simulations of Eq.\ (\ref{ODEs}); for different values of the number of states, $n$, the order parameter, $r$, is plotted as a function of the coupling parameter $\alpha$.  It appears that the critical exponent follows the $\gamma=1/2$ prediction of the heuristic approach to very good accuracy.  A critical exponent $\gamma=1/2$ was already established for $n=4$ in Ref.\ \cite{BierLisowskiNowak}.  We also obtained the plots for $n=6$ and $n=8$.  For these values we likewise found a critical exponent of 1/2.  For values $n \gtrapprox 10$ the numerical simulations become inaccurate, especially in the vicinity of the phase transition.  There are therefore insufficient data to verify that $3/(2 \sqrt{n})$ prefactor in Eq.\ (\ref{OrderParTheor}).
     
\subsection{The Temperature Dependence}

If transitions as in Eq.\ (\ref{TransRates}) are thermally activated, then the transition rate $k$ generally follows an Arrhenius dependence and features the temperature, $T$, in the denominator of the exponent, i.e.\ $k \propto \exp \left[ -b/T \right]$, where $b$ is positive \cite{MoorePhysChem}.  This means that the coupling parameter $\alpha$ in Eq.\ (\ref{TransRates}) should be replaced by $\alpha/T$ if we wish to include temperature dependence.  We can now see how the system behaves in the experimentally more likely scenario where the temperature is varied while coupling constants are fixed and constant.  Equation (\ref{amplitude}) has the form $r \propto \sqrt{\alpha - \alpha_c}$, where $\alpha_c$ denotes the critical value of $\alpha$. For a constant $\alpha = \alpha_0$ and a varying temperature $T$, this formula takes the form $r \propto \sqrt{\alpha_0/T - \alpha_0/T_c} \propto \sqrt{1/T - 1/T_c}$.  When $T$ is close to $T_c$, the latter expression is well approximated by $r \propto \sqrt{T_c - T}$.  This is the same mean-field temperature dependence of the order parameter that we discussed earlier in the context of the gas-liquid transition (cf.\ Fig.\ 1) and the onset of magnetization (cf.\ Fig.\ 2).    

\section{Conclusions and Discussion}

\subsection{The Phase Transition as the Number of States is Increased} 

The system of Fig.\ 3a with $n=3$ has already been a starting point for much research \cite{wood1,wood2,wood3,wood4}.  What we observe for $n=3$ is a critical exponent of $1/4$ (see Fig.\ 4).  A bifurcation with a critical exponent of 1/4 occurs in $\dot x = x(\mu - x^4)$ as the system crosses $\mu=0$.  For our system in case of $n=3$, the nonlinear system of three coupled first-order equations is highly coupled; each of the three equations contains nonlinear terms that contain all of the three dependent variables.  It should furthermore be pointed out that this system of three coupled first-order ordinary differential equations actually readily reduces to second order.  This is because of the fixed and normalized total population, i.e.\ $p_1 + p_2 + p_3 = 1$.  

For $n=4$ each state is coupled to two of the four others states.  In Ref.\ \cite{BierLisowskiNowak} the system of ODEs is numerically solved and, in addition, stochastic simulations are performed and a heuristic derivation is presented.  The critical exponent comes out to be 1/2.  For $n=4$, we still face a highly structured dynamical system and it is remarkable that the statistical approach and associated the mean-field prediction apparently already apply.  

In the system of Fig.\ 3a, each state ``interacts'' with its two neighbor states.  So for general $n$, each oscillator effectively connects with a fraction of about $2/n$ of the entire population.  It appears that for $n=4$, the fraction of 1/2 is sufficiently high to warrant a mean-field approach.  As $n$ is increased, we expect the legitimacy of the mean-field approach to break down.  In Ref.\ \cite{BierLisowskiNowak} the continuum limit, $n \rightarrow \infty$, of the system in Fig.\ 3a is investigated.  In that limit the flow of probability density around the cycle in Fig.\ 3a is described by a PDE that, after some manipulation, appears equivalent to a Burgers' Equation \cite{AtleeJackson}.  It turns out that for $n \rightarrow \infty$, a phase transition no longer occurs as $\alpha$ is varied.  The results presented in section 2 of this article are consistent with this observation: in the $n \rightarrow \infty$ limit, the phase transtion is pushed to the $\alpha \rightarrow \infty$ limit.  Going to $n=10$ with our numerical simulations, we did not observe a change of the mean-field exponent of 1/2.   All in all, both the simple heuristic approach and the full simulation of the ODEs show that the critical exponent of the phase transition keeps the mean-field value of 1/2, but that the phase transition occurs for ever higher values of the coupling parameter, $\alpha$, as the number of states, $n$, is increased.  We will come back to this in the last subsection.   

\subsection{Entropy and Dissipative Structures}

As was mentioned before, our system as depicted in Fig.\ 3 has irreversible transitions.  Unlike the systems discussed in sections 1A and 1B, it is not at equilibrium.  Our system produces entropy and in this subsection we will come to a quantitative assessment of the involved entropies.

One oscillator with a probability $p_k$ of being in state $k$, comes with an associated entropy of $S= - \sum_{k=1}^n p_k \log p_k$.  For the case of a homogeneous distribution, it is readily found that $S = \log n$.  If we have $p_k = (1+\varepsilon)/n$ for half of the $n$ states and $p_k = (1-\varepsilon)/n$ for the other half, then we find $S \approx \log n - \varepsilon^2/2$ after we use the approximation $\log (1 \pm \varepsilon) \approx \pm \varepsilon - \varepsilon^2/2$.  In other words, the nonhomogeneous distribution of oscillators over states leads to an entropy decrease.  Clustering decreases entropy. 

Consider again the ``bump'' in Fig.\ 3b.  If the probability in each state were $1/n$, then the flux from each state to the next would be $J_{i \rightarrow (i+1)} = p_i k_i = 1/n$.  However, when there is a  jump as in Fig.\ 3b,  the two fluxes adjacent to the jump carry nonzero exponents (cf.\ Eq.\ (\ref{TransRates})).  From state $i$ to $i+1$, the probability goes from $(1-\varepsilon)/n$ to $(1+\varepsilon)/n$.  This leads to $J_{i \rightarrow (i+1)} = (1/n) (1-\varepsilon) \exp \left[ 2 \alpha \varepsilon/n \right]$ and $J_{(i+1) \rightarrow (i+2)} = (1/n)(1+\varepsilon) \exp \left[ 2 \alpha \varepsilon /n \right]$.  Upon going from state $(i+m)$ to state $(i+m+1)$, there is a decrease from $(1+\varepsilon)/n$ to $(1-\varepsilon)/n$.  This leads to $J_{(i+m) \rightarrow (i+m+1)} = (1/n)(1+\varepsilon) \exp \left[ -2 \alpha \varepsilon / n \right]$ and $J_{(i+m+1) \rightarrow (i+m+2)} = (1/n)(1-\varepsilon) \exp \left[ -2 \alpha \varepsilon / n \right]$.  The remaining fluxes along the horizontal axis in Fig.\ 3b are $(1+\varepsilon)/n$ and $(1-\varepsilon)/n$ in the elevated and lowered part, respectively.  They average to $1/n$ as there are just as many elevated as lowered probabilities.  It is readily verified that the four fluxes adjacent to the upward and downward jump average to $(1/n) \cosh \left[ 2 \alpha \varepsilon/n \right]$.  For any nonzero real value of $x$, we have $\cosh x > 1$.  This means that having a ``bump'' leads to a higher throughput for the cycle in Fig.\ 3a and to a larger production of entropy. 

The phase transition from a homogeneous distribution to one with ``bumps'' constitutes a symmetry breaking and an establishment of an order.  However, this lower entropy structure leads to a larger throughput and a larger entropy production for the system as a whole.  We can thus view the bumps as a self-organized dissipative structure as described by Prigogine in the 1970s \cite{Prigogine}. 

The increase of four fluxes from an average of $1/n$ to an average of $(1/n) \cosh \left[ 2 \alpha \varepsilon/n \right]$ for every bump can help us understand why the phase transition is pushed out to $\alpha \rightarrow \infty$ for $n \rightarrow \infty$.  We have $|\varepsilon| < 1$.  So for $n \rightarrow \infty$, the argument $2 \alpha \varepsilon/n$ vanishes (leading to $(1/n) \cosh \left[ 2 \alpha \varepsilon/n \right] \rightarrow 1/n$) and the enhanced flux disappears,  {\em unless} $\alpha$ changes proportionally with $n$.     

The subject matter of this subsection can be the starting point for ample mathematical analysis.  There is, for instance, a large body of work on how the flux through an entire cycle as in Fig.\ 3a is affected if just a few transitions are speeded up \cite{Heinrich,Kacser}.  As Eq.\ (\ref{amplitude}) is a rough approximation already, it would be somewhat excessive to substantially elaborate in this direction.   

\subsection{Why Mean-Field Works for a 1D System}

It may at first seem surprising that a 1D system as in Fig.\ 3a gives rise to a value of the critical exponent that is  characteristic of the mean-field approximation.  After all, the 1D Ising model features no phase transition at all if the number of involved spins, $N$, is finite.  Only if the 1D Ising array has infinitely many spins is there a phase transition, but it occurs at the $T \rightarrow 0$ limit if $N \rightarrow \infty$. 

However, upon closer consideration our mean-field result makes sense.  With the system in Fig.\ 3a we face a large number of oscillators, $N_{tot}$, that is distributed over $n$ states.  For the mass action approach of Eq.\ (\ref{ODEs}) to apply, we need $N_{tot} \gg n$.  Let $N_i$ be the number of oscillators in state $i$.  We then have $p_i = N_i/N_{tot}$.  The rate of change of $N_i$ depends on the numbers $N_{i-2}$, $N_{i-1}$, $N_{i}$, and $N_{i+1}$.  In this way every individual oscillator is interacting with infinitely many other oscillators.  The legitimacy of the mean-field approach can thus be understood.     

The 1D Ising model can be evaluated analytically and the pertinent derivation is shown in many authoritative textbooks \cite{ChaikinLubensky,PlischkeBergersen,Salinas}.  A rigorous treatment shows that at finite temperature, the magnetization (which is taken as an order parameter in Ising models) is zero in the absence of an external field.  However, as the temperature $T$ goes to zero, a singularity is approached.  Complete alignment, i.e.\ $r=1$, occurs at $T=0$ \cite{PlischkeBergersen}.  Lev Landau gave a good intuitive explanation of this behavior \cite{Salinas}.  We will give an explanation in the same vein for our coupled oscillators, but before doing so it is instructive to reiterate Landau's account.

Let $-J$ and $+J$ be the energies for neighboring spins in the parallel and antiparallel alignment, respectively.  Next imagine a cyclic 1D array of $N$ spins all in parallel alignment.  We randomly pick two spins on this cycle.  There are $W=N(N-1)$ ways to make this choice.  The chosen spins divivide the cycle into two segments.  After flipping all the magnets in one of the segments, we have two magnetic domains.  The flipping requires $4J$ of free energy, but is associated with an entropy increase of $\Delta S = k_B \ln W \approx 2 k_B \ln N$.  For finite $T$ and $N \rightarrow \infty$, we always have $T \Delta S \gg 4 J$.  This means that no finite coupling energy is sufficient to overcome the thermal noise and give full alignment of all spins.  Only at $T=0$ is it possible to achieve $r>0$.

Going back to the system depicted in Fig.\ 3a, a similar line of reasoning is found to apply.  Assume we add one oscillator to the system with $N_{tot}$ oscillators.  The randomization over $n$ states is associated with an entropy of $S = k_B \ln n$.  This quantity obviously increases with $n$.  As was mentioned before, the energy that is driving the transitions features in the exponents of the transition rates $k_i$ (cf.\ Eq.\ (\ref{TransRates})).  If an oscillator is added to state $i$, there is a change in the quantities $\alpha (p_{i+2} - p_i)$ and $\alpha (p_{i} - p_{i-2})$ that are in the exponents of $k_{i+1}$ and $k_{i-1}$, respectively.  For an increasing number of states, one oscillator is coupled to an ever smaller fraction of the entire population of oscillators.  The changes in the coupling energy will therefore become ever more insignificant upon increase of $n$ when compared to $k_B T \ln n$, i.e.\ the free energy change due to entropy effects.  This explains why the value of the coupling constant $\alpha$ at which the phase transition occurs increases with $n$.  Equivalently, the value of $T$ at which the phase transition occurs decreases with $n$ and we have $T_c \rightarrow 0$ as $n \rightarrow \infty$.  

\subsection{Implications for Coupled Oscillators in the Wet Lab}  

We have seen and come to understand that the 2nd order phase transition with a critical exponent of 1/2 is very robust.  The mean-field value of the critical exponent arises because the mass action (Eq.\ (\ref{ODEs})) implies that each oscillator in the system is coupled to infinitely many other oscillators.  Coupling an oscillator to a smaller fraction of the total number of oscillators pushes the phase transition to a higher value of the coupling coefficient or, equivalently, to a smaller value of the temperature.  With this insight we expect that the critical exponent value of 1/2 will also persist if the transition rates in Fig.\ 3a are made to vary along the cycle or if other mathematical forms for the population dependencies of the the transition rates are tried.  

\begin{figure}
\centering{\resizebox{9 cm}{!}{\includegraphics{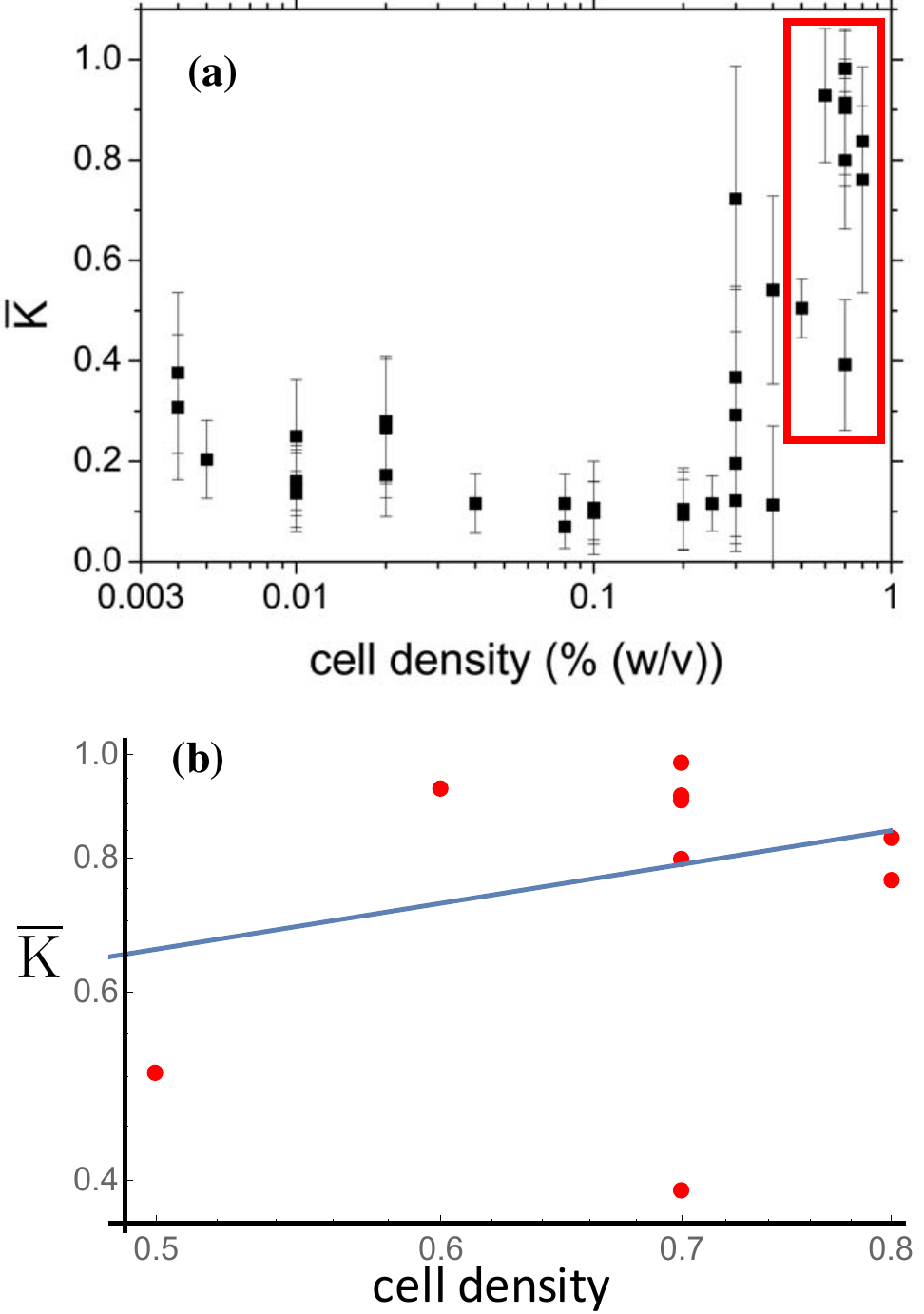}}}
\caption{(a) From Ref.\ \cite{Hauser}.  Yeast cells in a suspension synchronize their glycolytic oscillations, but do so more effectively if the density of the cells in the suspension is higher.  The horizontal axis (note the logarithmic scale) gives the density of the yeast cells in the suspension.  On the vertical axis $\bar{K}$ gives the resulting order parameter for the coupled oscillating cells (cf.\ Eq.\ (\ref{OrderParam})).  From a density of about 0.5\% on, we observe a rapid increase $\bar{K}$ and what appears like a phase transition.  (b) Taking the data points in the red rectangle and giving the vertical axis a logarithmic scale, we find for the best fitting straight line a slope of 0.54.  The margins of error are large and data points exhibit a wide spread, but this result appears consistent with the 1/2 that the Kuramoto model and our theory predict.}
\end{figure}

There is autocatalysis or product stimulation in the system depicted in Fig.\ 3a, i.e., the product of the $j \rightarrow j+1$ transition increases the rate of the $j \rightarrow j+1$ transition.  Product stimulation is a form of positive feedback and it is commonly the underlying driving force behind biochemical oscillations \cite{NovakTyson}.  In the system in Fig.\ 3a, the parameter $\alpha$ can be seen as a measure for the strength of the positive feedback.  Product stimulation is a key feature in the glycolytic oscillations that were already mentioned in Section 1D \cite{BiophysChem1996}. The product stimulation is twofold.  ATP binding and hydrolysis is the first step in this metabolic chain, but subsequently ATP is produced again at several steps in the chain.  Furthermore, early in the chain energy that is harvested from the breakdown of glucose is stored in NAD reduction: NAD$^+$ + H$^+$ + 2e$^-$ $\rightarrow$ NADH.  For the final step in the chain, the conversion from acetaldehyde to ethanol, the necessary energy is derived from the oxidation of NADH, i.e.\ the reverse reaction: NADH $\rightarrow$ NAD$^+$ + H$^+$ + 2e$^-$.  The glycolysis consists of about ten enzyme-mediated steps and because ATP and NADH are products as well as substrates in the chain, the chain can be seen as a cycle much like the one in Fig.\ 3a.  

In a suspension of oscillating yeast cells there is a synchronized oscillation inside every cell.  The number of oscillators inside each cell, $N_{in}$, is different for each cell.  The catalyzing enzymes process the substrate molecules one-by-one and it is substrate concentrations and the number of enzymes in each cell that ultimately determine an effective $N_{in}$ and a characteristic frequency for each cell. 

As was mentioned before, acetaldehyde can freely permeate the membrane of the yeast cell.  The oscillations are thus coupled through the shared acetaldehyde concentration.  It is obvious that for a suspension with a small density of yeast cells, the acetaldehyde concentration will be close to zero and not lead to any coupling.  Reference \cite{Hauser} models the suspension of oscillating yeast cells with a Kuramoto model, where the shared acetaldehyde concentration provides the coupling between the $M$ cells.  Figure 5a is from Ref.\ \cite{Hauser} and shows the order parameter as a function of the density.  The measurements are not very precise, but appear consistent with the phase transition and the critical exponent of 1/2 that are predicted by the Kuramoto model.  A model as in Fig.\ 3a, which also predicts a critical exponent of 1/2, may be more chemically realistic than the Kuramoto model as it explicitly includes positive feedbacks that underly the oscillations.

There is still a measure of controversy about the oscillations of yeast cells in a suspension.  Reference \cite{Hauser} reports the observation that individual cells still oscillate even when they are in a solution that is too dilute for coupling.  In Ref.\ \cite{QuorumSensing}, on the other hand, similar measurements were described and there it was found that cells in a diluted solution no longer oscillate.  The authors of Ref.\ \cite{QuorumSensing} describe how a suspension of cells ultimately uses the acetaldehyde concentration for ``quorum sensing.''  Only when a quorum is met, i.e.\ when the density is sufficiently high, do oscillations commence.  A Kuramoto model is in that case no longer appropriate.  Quorum-sensing can be consistent with a model as in Fig.\ 3a.  In a very dilute solution acetaldehyde effectively diffuses away and disappears from the cell as soon as it is formed.  The conversion to ethanol and the accompanying NADH consumption then no longer take place.  Instead of an NADH-NAD feedback loop, we would get NADH accumulation.  Upon increase of the cell density, the NADH-NAD feedback loop gets established.  Once the quorum is met, a mean-field approach applies and a phase transition with a critical exponent of 1/2 once again occurs.     

\section{Acknowledgments}

We are grateful to Katja Lindenberg, Hans Westerhoff, and Bartek Lisowski for very useful feedback on earlier versions of this manuscript.

\end{document}